\documentclass[12pt]{article}
\usepackage{amsmath}
\usepackage{amscd}
\usepackage{amssymb}
\usepackage{array}

\begin{document}
\rightline{CERN-TH/2002-346}
\rightline{UCLA/02/TEP/36}

\newcommand{\R}{\mathbb{R}}
\newcommand{\C}{\mathbb{C}}
\newcommand{\Z}{\mathbb{Z}}
\newcommand{\Hb}{\mathbb{H}}

\newcommand{\rE}{\mathrm{E}}
\newcommand{\rSp}{\mathrm{Sp}}
\newcommand{\rSO}{\mathrm{SO}}
\newcommand{\rSL}{\mathrm{SL}}
\newcommand{\rSU}{\mathrm{SU}}
\newcommand{\rUSp}{\mathrm{USp}}
\newcommand{\rU}{\mathrm{U}}
\newcommand{\rF}{\mathrm{F}}
\newcommand{\rGL}{\mathrm{GL}}
\newcommand{\rG}{\mathrm{G}}
\newcommand{\rK}{\mathrm{K}}

\newcommand{\fgl}{\mathfrak{gl}}
\newcommand{\fu}{\mathfrak{u}}
\newcommand{\fsl}{\mathfrak{sl}}
\newcommand{\fsp}{\mathfrak{sp}}
\newcommand{\fusp}{\mathfrak{usp}}
\newcommand{\fsu}{\mathfrak{su}}
\newcommand{\fp}{\mathfrak{p}}
\newcommand{\fso}{\mathfrak{so}}
\newcommand{\fl}{\mathfrak{l}}
\newcommand{\fg}{\mathfrak{g}}
\newcommand{\fr}{\mathfrak{r}}
\newcommand{\fe}{\mathfrak{e}}
\newcommand{\ft}{\mathfrak{t}}

\vskip 1cm

  \centerline{\LARGE \bf Gauged extended supergravity without }

  \bigskip

   \centerline{\LARGE \bf   cosmological constant: no-scale structure }
    \bigskip

   \centerline{\LARGE \bf   and supersymmetry breaking.}

 \vskip 1.5cm
\centerline{L. Andrianopoli$^\flat$, R. D'Auria$^\sharp$,  S.
Ferrara$^{\flat \star}$ and M. A. Lled\'o$^\sharp$.}

\vskip 1cm

\centerline{\it $^\flat$ CERN, Theory Division, CH 1211 Geneva 23,
Switzerland.}

\bigskip

\centerline{\it $^\sharp$ Dipartimento di Fisica, Politecnico di
Torino,} \centerline{\it Corso Duca degli Abruzzi 24, I-10129
Torino, Italy  and  } \centerline{\it   INFN, Sezione di Torino,
Italy. }

\bigskip

\centerline{$^\star$\it Department of Physics and Astronomy,} \centerline{\it University of
California,
Los Angeles. Los Angeles, USA, and} \centerline{\it INFN, Laboratori
Nazionali di
Frascati, Italy}

\vskip 1.5cm

\begin{abstract}
We consider the interplay of duality symmetries and gauged isometries of supergravity models giving
$N$-extended, spontaneously broken supergravity with a no-scale structure. Some examples motivated by
 superstring and M-theory compactifications are described.
\end{abstract}

\vskip 1.5cm

Invited paper to appear in the review section of the
``International Journal of Modern Physics A".

 \vfill\eject

\section{Introduction}

An important feature of a generic supergravity theory is the
possibility of undergoing spontaneous supersymmetry breaking
without a cosmological constant. By studying the universal
coupling of a Goldstone fermion to supergravity, one can see that
in a spontaneously broken supergravity theory the contributions
to the vacuum energy could in principle cancel \cite{dz}. The
first concrete example, based on a field theory lagrangian, was
given by Polony \cite{po,cjsfgvn}. He considered $N=1$
supergravity coupled to a single chiral multiplet with canonical
kinetic term and linear superpotential and showed that it is
possible to fine tune the parameters ($\alpha$ and $\beta$) of the
superpotential $W=\alpha z+\beta$ in such a way that the potential
stabilizes the scalar fields with vanishing vacuum energy.
 The scalar field masses satisfy the sum rule $m_A^2+m_B^2=4m^2_{3/2}$ \cite{cjsfgvn}.
 Polony type superpotentials were used in the first phenomenological studies of broken
supergravity. They generate the soft breaking terms of the observable sector of standard
(electroweak and strong) interactions in the supersymmetric extension of the standard model
and of grand unified theories (For a review, see Ref. \cite{ni}).

 The Polony classical potential is rather unnatural because it requires an ad hoc
superpotential. Subsequent studies of the superHiggs sector  of supergravity models lead to
the introduction of a more appealing class of theories, the so called no-scale supergravities
\cite{cfkn,elnt}. In these models,  the vanishing vacuum energy of the classical potential is
obtained without stabilizing the scalar superpartner of the Goldstino.  Instead, there is an
exact cancellation, prior to minimization, of the positive Goldstino contribution against the
negative gravitino contribution to the vacuum energy  without the need of fine-tuning the
parameters.
 The no-scale structure of these models poses further constraints on the soft-breaking terms
which enter in the phenomenological Lagrangians \cite{ln}.

 The first construction of an extended supergravity exhibiting a no-scale structure was in the
context of  $N=2$ supergravity coupled to abelian vector multiplets in presence of a
Fayet-Iliopoulos term \cite{ckpdfwg}. For a certain choice of the geometry of the scalar
manifold, a spontaneous breaking of $N=2$ to $N=0$ with flat potential takes place.  Later,
models with $N=2$ supersymmetry partially broken to $N=1$ were found. In these models the
vector multiplets gauge  particular isometries of the quaternionic variety pertaining to the
hypermultiplets.  The breaking of  supersymmetry with naturally vanishing vacuum energy was
achieved by gauging  two traslational isometries of the quaternionic manifold \cite{cgp,fgp}.
One unbroken supersymmetry required a relation between the  gauge coupling constants of the
two translational isometries. To the  $n_t$ translational isometries of the quaternionic
manifold correspond  $n_t$ axion fields $b^i$ transforming by a shift. We can express the
shift corresponding to the gauged isometries as
 $$b^i(x)\longrightarrow b^i(x) +g_1^i\xi^1(x)+g_2^i\xi^2(x),$$
 so the covariant derivatives are
 $$\mathcal{D}_\mu b^i=\partial_\mu b^i-g_1^iA_\mu^1(x)-g_2^iA_\mu^2(x).$$

 The simplest model \cite{cgp,fgp} is based on the quaternionic manifold of quaternionic
dimension $n_H=1$
 $$\frac{\rSO(1,4)}{\rSO(4)}\simeq\frac{\rUSp(2,2)}{\rUSp(2)\times \rUSp(2)}.$$
It has three translational isometries, $i=1,\dots 3$. By choosing
a gauging such that $g_1^1=g$, $g_2^2=g'$ and zero otherwise,
one unbroken supersymmetry implies that $|g|=|g'|$.

It was later shown \cite{fgpt,zi} that it is possible to couple this $N=2$ hidden sector to
observable matter for a suitable choice of the vector and hypermultiplet geometry and for
appropriate gauge groups.

The no-scale structure for $N>2$ extended supergravity is encountered in the context of eleven
dimensional supergravity with Scherk-Schwarz generalized dimensional  reduction. This produces
spontaneously broken supergravity theories in four dimensions \cite{ss,css}.
  The four dimensional interpretation of these theories  \cite{adfl2} is an $N=8$ gauged
supergravity whose gauge algebra (a ``flat" algebra according to Ref. \cite{ss}) is a 28
dimensional Lie subalgebra of $\rE_{7,7}$ obtained in the following way: Consider the
decomposition
 $$\begin{CD}\fe_{7,7}@>>\fe_{6,6}+\fso(1,1)>\fe_{6,6}+\fso(1,1)+\mathbf{27}^++\mathbf{27}^-
 \end{CD},$$
 Then the flat subalgebra is the semidirect sum of a factor $\fu(1)$ in the Cartan subalgebra
of $\fusp(8)$ (maximal compact subalgebra of $\fe_{6,6}$) with the 27 translational subalgebra
$\mathbf{27}^-$.The commutation rules are
 \begin{eqnarray*}&&[X_\Lambda, X_0]=f_{\Lambda 0}^{\;\Sigma}
X_\Sigma,\\&&[X_\Lambda,X_\Sigma]=0\qquad \Lambda=1,\dots
27,\end{eqnarray*}
with $f_{\Lambda 0}^{\;\Sigma}
=M_\Lambda^\Sigma$ in the CSA of $\fusp(8)$.

The 27 axions $a^\Lambda$ in $\rE_{7,7}/\rSU(8)$ transform
 under the gauge algebra as follows:
$$\delta a^\Lambda=M^\Lambda_\Sigma\xi^\Sigma+\xi^0M_\Sigma^\Lambda a^\Sigma,$$
and their covariant derivatives, in terms of the gauge fields $B_\mu$ and $Z^\Sigma_\mu$, are
$$\mathcal{D}_\mu a^\Lambda=\partial_\mu a^\Lambda- M_\Sigma^\Lambda a^\Sigma B_\mu
-M_\Sigma^\Lambda Z^\Sigma_\mu.$$
The gauge fields transform as
\begin{eqnarray*}
\delta Z_\mu^\Lambda&=&\partial_\mu\xi^\Lambda+\xi^0M_\Sigma^\Lambda Z^\Sigma_\mu-\xi^\Sigma
M_\Sigma^\Lambda B_\mu\\
\delta B_\mu&=&\partial_\mu\xi^0
\end{eqnarray*}
With respect to $\rUSp(8)$ the representation {\bf 27} of $\rE_{7,7}$ is the two fold
antisymmetric $\Omega$-traceless representation, so we can write $\Lambda\rightarrow
(a_1,a_2)$ and
$$M^\Lambda_\Sigma \rightarrow
M^{[a_1}_{[b_1}\delta_{b_2]}^{a_2]}-\Omega\!\!-\!\!\mathrm{traces}$$ where $M_{b_1}^{a_1}$
turns out to be the gravitino U(1)-charge matrix,
\begin{eqnarray}\begin{pmatrix}m_1\epsilon&0&0&0\\0&m_2\epsilon&0&0\\0&0&m_3\epsilon&0\\0&0&0&m_4
\epsilon
\end{pmatrix}\label{matrix}\end{eqnarray}
with $\epsilon=\begin{pmatrix}0&1\\-1&0\end{pmatrix}$.
The gravitino  mass matrix is the symmetric 8$\times$8 matrix
$$S_{ab}=e^{-3\phi}(M\Omega)_{ab},$$
where $\phi$ is the radion field \cite{adfl2,adfl3}. This is the constant term  in the fifth
component of the $\rUSp(8)$ connection of $\rE_{6,6}/\rUSp(8)$ upon generalized dimensional
reduction \cite{svn}.

The most recent example of no-scale extended supergravity is the
$N=4$ spontaneously broken theory \cite{adfl3,tz} which is the low
energy effective action for type IIB superstrings on type IIB
orientifolds in presence of D3-branes and with three-form fluxes
turned on \cite{fp}. In presence of $n$ D3-branes  this theory
corresponds to a gauged supergravity with gauge group the direct
product $T_{12}\times \rU(n)$ which are a particular set of
isometries of the sigma model
$\rSO(6,6+n^2)/\rSO(6)\times\rSO(6+n^2)$. The latter is a sigma
model of an $N=4$ supergravity theory coupled to $6+n^2$ vector
multiplets. In the superstring interpretation, six of the vector
multiplets come from the supergravity fields on the bulk and the
rest comes from a non abelian D3-brane Born-Infeld action coupled
to supergravity. The twelve bulk vectors gauge the $T_{12}$ factor
and the gauge vectors living on the brane gauge the $\rU(n)$
Yang-Mills  group. The full action has been recently constructed
\cite{dfv}. It is a no-scale $N=4$ supergravity with four
arbitrary gravitino masses and its moduli space is a product of
three non compact projective spaces
$\rU(1,1+n)/\rU(1)\times\rU(n)$ \cite{fpo}.  If we formally
integrate out step by step the three massive gravitino multiplets
(in this process $N=4\rightarrow N=3\rightarrow N=2\rightarrow
N=1$) we end up with an $N=1$ no-scale supergravity with a
particular simple form  which falls in the class of no-scale
models studied in the literature \cite{ln}.

 \bigskip

 The paper is organized as follows. In Section 2 we review the scalar potential in
$N$-extended supergravity and outline the properties of no-scale
supergravities regardless of the specific matter content and of
the number of supersymmetries. In Section 3 we formulate the $N=8$
and $N=4$ spontaneously broken theories discussed so far in the
context of no-scale gauged supergravities. In Section 4 we briefly
review the $N=1$ no scale models and consider the $N=1$ type IIB
orientifold model in this framework.

 \section{Scalar potential in N-extended supergravity: vacua without cosmological constant}

 We consider an $N$-extended supergravity theory in $D=4$. We will denote by $\psi_{\mu A}, \;
A=1,\dots N$ the spin 3/2 gravitino fields and by $\lambda^I$ the
spin $1/2$ fields   (the spinor indices are not shown explicitly).
They are all taken to be left handed, and the right handed
counterparts are denoted by $\psi_{\mu }^A$ and $\lambda_I$. The
scalar fields will be denoted by $q^u$, and are coordinates on a
Riemann manifold $\mathcal{M}$. Supersymmetry requires that
$\mathcal{M}$  has a restricted holonomy group $H=H_R\times H_M$,
with $H_R$ being $\rU(N)$ or $\rSU(N)$ ($\rU(N)$ being the
R-symmetry group) and $H_M$ varying according to the different
matter multiplet species. It also requires that on $\mathcal{M}$
there is an $H_R$-bundle with a connection whose curvature  is
related to the geometric structure of  $\mathcal{M}$
\cite{abcdffm}.

For $N=1$  supergravity coupled to $n$ chiral multiplets we have a
K\"ahler-Hodge manifold of complex dimension $n$, with
$H_M=\rSU(n)$ and $H_R=\rU(1)$. On $\mathcal{M}$ there is a U(1)
bundle whose Chern class is equal to the K\"ahler class.

For $N=2$ coupled to $n$ vector multiplets we have a special
K\"ahler-Hodge manifold. If we have $n_h$
 hypermultiplets, then $\mathcal{M}$ is a quaternionic manifold of quaternionic dimension $n_h$. The
holonomy is $H=H_R\times H_M$ with $H_R=\rSU(2)$ and
$H_M=\rUSp(2n_h)$. On $\mathcal{M}$ there is an $\rSU(2)$-bundle
with curvature equal to the triplet of hyperK\"ahler forms on
$\mathcal{M}$.

For $N>2$ the manifolds of the scalars are maximally symmetric spaces $\mathcal{M}=G/H$ with
$H=H_R+H_M$. Then there is also an $H_R$-bundle on $\mathcal{M}$ whose connection is the $H_R$ part
of the spin connection.

For $N=3$ with $n_v$ vector multiplets $H_M=\rSU(n_v)$ and $H_R=\rU(3)$. For $N\geq 4$ the
supergravity multiplet itself contains scalars. For $N=4$ with $n_v$ vector multiplets
$H_M=\rSO(n_v)$ and $H_R=\rSU(4)\times \rU(1)$. For $N>4$ there are no matter multiplets and
$H=H_R=\rU(N)$ except for $N=8$ where $H=\rSU(8)$ \cite{adf}.

The above considerations imply that the covariant derivative of
the supersymmetry parameter, $\mathcal{D}_\mu\epsilon_A$,
contains, in presence of scalar fields, an $H_R$ connection in
addition to the spacetime spin connection.

\bigskip

 The supersymmetry variations of the fermions in a generic supergravity theory can be expressed as
\cite{df}
 \begin{eqnarray} \delta\psi_{A\mu}=\mathcal{D}_\mu\epsilon_A+\frac 1 2
S_{AB}\gamma_\mu\epsilon^B+\cdots\noindent\\
 \delta\lambda^I=iP_{\mu A}^I\gamma^\mu\epsilon^A+
N^{IA}\epsilon_A+\cdots,\label{varfer}\end{eqnarray}
 where $S_{AB}=S_{BA}$,  and $N^{IA}$ are sections of  $H_R$ bundles on $\mathcal{M}$ which depend on
 the specific model under consideration. The dots stand for terms which contain vector fields.
 $P_{\mu A}^Idx^\mu$ is pullback  into spacetime of $P_{uA}^Idq^u$, the vielbein one-form on
$\mathcal{M}$, so $$P_{\mu A}^I=P_{uA}^I\partial_\mu q^u.$$ The variation of the scalars is then
given by
 $$\delta q^uP_{uA}^I=\bar \lambda^I\epsilon_A.$$

 The supergravity lagrangian contains the following terms
 $$\frac{1}{\sqrt{-g}}\mathcal{L}=\cdots
+S_{AB}\bar\psi_\mu^A\sigma^{\mu\nu}\psi_\nu^B+iN^{IA}\bar\lambda_I\gamma^\mu\psi_{\mu A} +M^{IJ}
\bar\lambda_I\lambda_J \quad+\mathrm{c.c.}\! -\!V(q)$$
 where $M^{IJ}$ is the mass matrix of the spin $1/2$ fields and $V(q)$ is the potential of the scalar
fields. The potential must be such that the supersymmetry variation of all these terms cancel. This
implies \cite{fm,cgp2}
 \begin{eqnarray}\delta_B^AV(q)=-3 S^{AC}S_{BC}+N^{IA} N_{IB}\label{potential}\\
 \frac{\partial V}{\partial q^u}P_{uA}^I=2iN^{IB}S_{BA} +2M^{IJ} N_{JA},\nonumber\end{eqnarray}
 where $N_{IA}=(N^{IA})^*$ and $S^{AC}=S_{AC}^*$.

 Flat space requires that on the extremes ${\partial V}/{\partial q^u}=0$ the potential  vanishes, so
 $$3\sum_CS^{AC}S_{CA}=\sum_IN^{IA} N_{IA},\qquad \forall A.$$ The first term in the potential
(\ref{potential}) is the square of the gravitino mass matrix. It is hermitian, so it can be
diagonalized
by a unitary transformation. Assume that it is already diagonal, then the eigenvalue in the entry
$(A_0,A_0)$ is non zero if and only if   $N^{IA_0}\neq 0$ for some $I$. On the other hand, if the
gravitino mass matrix vanishes then $N^{IA}$ must be zero.

 For no-scale models, there is a subset of fields $\lambda^{I'}$ for which
 \begin{equation}3\sum_CS^{AC}S_{CA}=\sum_{I'}N^{I'A} N_{I'A},\qquad \forall
A\label{cancel}\end{equation}
in all $\mathcal{M}$. This implies that the potential is given by
 $$V(q)=\sum_{I\neq I'}N^{IA} N_{IA},$$
 and it is manifestly positive definite. Zero vacuum energy on a point of $\mathcal{M}$ implies that
$N^{IA}=0$, $I\neq I'$  at that point. This happens independently of the number of unbroken
supersymmetries, which is controlled by $N^{I'A}$ (gravitino mass matrix).

 In $N$ extended supergravities, the axion couplings  to the gauge fields
 $$\mathcal{D}_\mu a^i=\partial_\mu a^i-g_\Lambda^iA_\mu^\Lambda$$
 are related to the gravitino mass matrix $S_{AB}$ through the existence, for each pair of indices
$i,\Lambda$ of a section $X^\Lambda_{i,AB}$ of an $H_R$ bundle over $\mathcal{M}$ such that
 $$S_{AB}=g_\Lambda^iX_{i,AB}^\Lambda, \qquad X_{i,AB}^\Lambda=X_{i,BA}^\Lambda.$$
 In the next section we will give the particular form of $S_{AB}$ in $N=8$ and $N=4$.

 \section{No-scale $N=8$ and $N=4$ theories}

 \subsection{$N=8$ Scherk-Schwarz spontaneously broken supergravity}

 In $N=8$ spontaneously broken supergravity \`a la Scherk-Schwarz, the $R$-symmetry that is manifest
is
$\rUSp(8)\subset \rSU(8)$. The spin 3/2 gravitinos are in the fundamental representation of $\rUSp(8)$
({\bf
8}), while the spin 1/2 fermions are in the {\bf 8} and {\bf 48} (threefold $\Omega$-traceless
antisymmetric representation). We will denote them as $\psi_{\mu a}, \lambda_a$ and $\lambda_{abc}$.

 From a dimensional reduction point of view, the scalar potential is originated by the five
dimensional
$\sigma$-model kinetic energy term
 $$\sqrt{-g}g^{\mu\nu}P_\mu^{abcd}P_{\nu abcd},\qquad \mathrm{for} \quad\mu=\nu=5, $$
 where $P_\mu^{abcd}$ is the pullback on spacetime of the  vielbein one form of the coset
$\rE_{6,6}/\rUSp(8)$.

 From the generalized dimensional reduction, the four dimensional scalar potential is
\begin{equation}V=\frac{1}{8}e^{-6\phi}P_5^{abcd}P_{5 abcd},\label{potred}\end{equation}
where $\phi$ is the radion field. This term would not appear in a standard dimensional reduction.

The five dimensional supersymmetry variations are
\begin{eqnarray*} \delta\psi_{\mu a}=\mathcal{D}_\mu\epsilon_a+\cdots\\\delta\lambda_{abc}=P_{\mu
abcd}\gamma^\mu\epsilon^d+\cdots.\end{eqnarray*}
We denote by $Q_{\mu ab}$ the $\rUSp(8)$ connection in five dimensions.
The functions $S_{AB}$ and $N^{IA}$ of the previous section (\ref{varfer}) are then
$$S_{ab}=\frac{1}{\sqrt3}e^{-3\phi} Q_{5 ab}, \quad N^{ab}=e^{-3\phi}Q_5^{ab}, \quad
N^{abcd}=e^{-3\phi}
 P_5^{abcd}$$
(the indices can be raised or lowered with the antisymmetric metric $\Omega_{ab}$). $P_5^{abcd}$ 
satisfies the identity \cite{svn}
$$P_5^{abcd}P_{5\,ebcd}=\frac{1}{8}\delta^a_eP_5^{fbcd}P_{5\, fbcd},$$
which is crucial to have (\ref{potential})

In the computation of the scalar potential using (\ref{potential}) there is an exact cancellation
between
the gravitino and the spin 1/2 fermions in the {\bf 8} as in (\ref{cancel}),
$$3|S_{ab}|^2=|N^{ab}|^2,$$
so that $$V=\frac{1}{8}|N^{abcd}|^2.$$
This explains formula (\ref{potred}) from a four dimensional point of view. Note that at a linearized
level (near the origin of the coset, where the exponential coordinates  $\phi^{abcd}$ are small),
\begin{eqnarray*}&&P_5^{abcd}= M^{[a}_{a'}\phi^{a'bcd]}\quad - \quad \Omega\!\!-\!\!\mathrm{traces}
+\mathcal{O}(\phi^{abcd})^2,\\&&Q_{5ab}\Omega^{ba'}= M_a^{a'}+ \mathcal{O}(\phi^{abcd}),\end{eqnarray*}
where $M_a^{a'}$ was given in (\ref{matrix}).

The vacua with zero potential correspond to $P_5^{abcd}=0$, while
the supersymmetry breaking depends on the vanishing eigenvalues of
the matrix  $Q_{5ab}$. When all the eignevalues $m_i$ of
(\ref{matrix}) are different from zero, the requirement
$P_5^{abcd}=0$ determines all but two coordinates which are the
two scalars which are neutral with respect to the CSA of
$\rUSp(8)$.  Together with the radion, they are the flat
directions of the potential. There are three additional massless
scalars, the three axions in the 27 of $\rUSp(8)$ which are
neutral under the CSA. All together, they form the moduli space of
the Scherk-Schwarz compactification and they parameterize the
coset $$\bigl(\frac{\rSU(1,1)}{\rU(1)}\bigr)^3.$$

If some eigenvalues $m_i$ of $M$ vanish, we have some unbroken
supersymmetries, and the moduli space of the solution is bigger.
If three masses are set to zero, then the equation $P_5^{abcd}=0$
leaves 14 coordinates undetermined, which parameterize
$$\frac{\rSU^*(6)}{\rUSp(6)},$$
 By adding the radion and 15 axions this space enlarges to $$\frac{\rSO^*(12)}{\rU(6)},$$ which is
the
moduli space of the $N=6$ unbroken supergravity.

 Similar reasoning can be used for the cases with two eigenvalues equal to zero ($N=4$) and only one
eigenvalue equal to zero ($N=2$). We observe that in all models the spin 1/2 fermions which cancel
the negative spin 3/2 contribution to the potential are precisely the Goldstino fermions. They are in
the {\bf 8} of $\rUSp(8)$ for  the $N=8$ model of section 3.1 and in the $\bar{\mathbf{4}}$ of
$\rSU(4)$ for the $N=4$ model of this section. When all supersymmetries are broken these fermions
disappear from the spectrum to give mass to the gravitino. If some supersymmetry remains unbroken
then these fermions are strictly massless.

\subsection{$N=4$ supergravity and type IIB orientifolds}

We now consider no-scale $N=4$  spontaneously broken supergravity. This theory is the low energy
limit of
type IIB 10 dimensional supergravity compactified on orientifolds in presence of three form fluxes
and $n$
D3 branes with non commutative coordinates \cite{fp,kst}.

The six $N=4$ vector multiplets coming from the bulk lagrangian
contain 36 scalars,  21 of which are the metric deformation of
the 6-torus $T^6$ $g_{IJ}$, $I,J=1,\dots 6$, and 15 scalars coming
from the four form gauge field $C_{\mu\nu\rho\lambda}$, whose
components along the 6-torus are dual to a two form
$$B^{IJ}=^*\!\!C^{IJ}, \qquad I,J=1,\dots 6.$$ Turning on the
three form fluxes corresponds in the effective theory to gauge
particular isometries of the coset
$\rSO(6,6)/\rSO(6)\times\rSO(6)$ \cite{adfl3}. More explicitly,
the gauge isometries are twelve of the fifteen translational
isometries in the graded decomposition \cite{adfl3}
$$\fso(6,6)=\fsl(6)+\fso(1,1) +{\mathbf{15}^+}+{\mathbf{15}^-}.$$

In the case when  Yang-Mills $N=4$ multiplets are added (describing the D3 brane degrees of freedom),
the
gauge group is $T_{12}\times \rU(n)$. This theory gives rise to a no-scale supergravity with four
arbitrary parameters for the gravitino masses \cite{tz}.

The SU(4) (R-symmetry) representations of the bulk fermions are as follows:

\noindent spin 3/2 (gravitinos)  in the {\bf 4},

\noindent spin 1/2 (dilatinos) in the $\bar{\mathbf{4}}$,

\noindent spin 1/2 (gauginos, from the 6 vector multiplets) in the {\bf 20}+$\bar{\mathbf{4}}$.

The fermions in the brane (gauginos) form $n^2=\dim\rU(n)$ copies of the representation {\bf 4} of SU(4).

Computing the potential (\ref{potential}), the subset of fields $\lambda^{I'}$ (\ref{cancel}) are the
bulk
gauginos  in the $\bar{\mathbf{4}}$. The condition for vanishing potential \cite{dfv}
is then $N^{IA}=0$ for $I\neq I'$. For the bulk fermions it fixes the complex dilaton, 18 radial
moduli
and 12 axions. For the brane gauginos it fixes all the scalars but the ones in the CSA of $\rU(n)$.

\section{$N=1$ no-scale supergravities}
Supergravity theories with a positive definite potential have a
particular convenient set up in the context of $N=1$ supergravity.
$N=1$ theories can be obtained from an $N$-extended supergravity
which is spontaneously broken to $N=1$ and then integrating out
the massive modes. This will be still true if $N=1$ is itself
spontaneously broken, provided the mass of the $N=1$ gravitino is
much smaller than the other gravitino masses.

To compute the contribution to the scalar potential of the chiral
multiplet sector of an $N=1$ theory it is convenient to introduce
some auxiliary fields \cite{fvn}. We denote by $u$ the auxiliary
field associated to the gravity multiplet and by $h^i$ the ones
associated to the chiral multiplets. Let $K$ be the K\"ahler
potential and  $W$ be the superpotential. We introduce the
function $$\frac \Phi 3=e^{-\frac K 3}.$$ Then, the scalar
potential is given by \cite{cfgvp}
\begin{eqnarray}-\bigl(\frac \Phi 3 \bigr)^2V=-\frac 1 9 \Phi|u|^2-\Phi_{i\bar j}h^ih^{\bar j}
+W_ih^i+\bar W_{\bar i}h^{\bar i} +\nonumber \\ \frac 1 3
u^*(3\bar W-\Phi_ih^i) + \frac 1 3 u(3 W-\Phi_{\bar i}h^{\bar
i}).\label{potentialaux}\end{eqnarray} (the derivatives with
respect to $z^i$ and $\bar z^{\bar i}$ are denoted by subindices
$i$ and $\bar i$). The standard potential is easily obtained by
making the field redefinition $$\tilde u=u-K_ih^i$$ so that
$$-\bigl(\frac \Phi 3 \bigr)^2V=-\frac 1 9 \Phi|\tilde u|^2+ \frac
1 3\Phi K_{i\bar j}h^ih^{\bar j}+ W{\tilde u} +\bar W{\tilde u}^*
+(K_i W+W_i)h^i+(K_{\bar i}\bar W+\bar W_{\bar i})h^{\bar i}.$$
Eliminating the auxiliary fields we get \cite{bw}
$$V=e^{K}\bigl[K^{i\bar j}\mathcal{D}_iW \mathcal{D}_{\bar j}\bar
W- 3 |W|^2],$$ where $\mathcal{D}_iW =\partial_iW+K_iW$.

The simplest example of no-scale supergravity is given by a
$\C\mathrm{P}^{n+1}$ $\sigma$-model \cite{ln} for which
$$\Phi=t+\bar t -\sum_Ac_A c_{\bar A}, \qquad A=1,\dots n$$ and an
arbitrary superpotential $W(c^A)$. From (\ref{potentialaux}),
since $\Phi_{t\bar t}=0$, then the variation with respect to $h^t$
implies $u=0$, and then the potential reduces to $$-\bigl(\frac
\Phi 3 \bigr)^2V= \sum_A(h^A h^{\bar A}  +h^AW_A+  h^{\bar
A}W_{\bar A}).$$ Then $$V=e^{\frac {2}{3}K}|\frac{\partial
W}{\partial c_A}|^2$$ so the extremes with vanishing vacuum energy
occur for ${\partial W}/{\partial c_A}=0$. In this example
 the gravitino mass contribution to the potential $3e^K|W|^2$ is canceled by the $\chi^t$ fermion
contribution.

 The crucial point here is that the matrix $\Phi_{i\bar j}$ has determinant zero and  rank $n$
\cite{bcf}. Such a situation generalizes to a class of models of
the following type $$\Phi=\prod _{r=1}^m(t_r+\bar t_r -\sum_A
c_{rA} c_{r\bar A})^\frac 1 m, \qquad A=1,\dots n$$ and $W$ a
function only of $c_{rA}$. This expression corresponds to the
K\"ahler potential of the product of $m$ spaces
$\C\mathrm{P}^{n+1}$ each with curvature $3/m$.

The function $\Phi$ is homogeneous of degree 1 in the variables $x_r=t_r+\bar t_r -
\sum_A c_{rA} c_{r\bar A}$. This implies that the matrix of second derivatives has a null vector
$$\sum_s\frac{\partial^2\Phi}{\partial x^r\partial x^s}x^s=0,$$ and from the form of the potential
this
implies that $u=0$. Then the potential becomes
$$-\bigl(\frac \Phi 3 \bigr)^2V= -\sum_{rs}\Phi_{rs}{\tilde h}^{t_r}{\tilde h}^{\bar t_s} +h^{c_{rA}}
h^{\bar c_{r\bar A}}\Phi_r+ h^{c_{rA}}W_{c_{rA}}+\bar h^{\bar c_{r\bar A}}\bar W_{c_{rA}},$$
where
$${\tilde h}^{t_r}={ h}^{t_r}-h^{c_{rA}}\bar c_{r\bar A}.$$

Eliminating the auxiliary fields one finds
$$V=-e^K\sum_r\frac{1}{K_r}|\frac{\partial W}{\partial c_{rA}}|^2=
e^K\sum_rK^{c_{rA}\bar c_{s\bar B}}\frac{\partial W}{\partial
c_{rA}} \frac{\partial \bar W}{\partial \bar c_{s\bar B}},$$ where
we have used the inverse of the K\"ahler metric $$K^{c_{rA}\bar
c_{s\bar B}}=-\delta_{rA \, s\bar B}\frac 1{K_{r}}.$$

\bigskip

An example of the above situation is realized in type IIB
orientifold with fluxes if one breaks $N=4$ to $N=1$ \cite{fpo}.
In this case $m=3$, $c^{rA}$ are the brane coordinates in the
adjoint of $\rU(n)$, and the superpotential is
$$W(c_{rA})=f+g^{ABC}c_{1A}c_{2B}c_{3C},$$ where $f$ is the
constant flux that breaks $N=1$ to $N=0$ and $g^{ABC}$ are the
structure constants of $\rSU(n)$.

\section*{Acknowledgements}

 M. A. Ll. would like to thank the Department of Physics and Astronomy of UCLA for its
kind hospitality during the  completion of this work. Work
supported in part by the European Community's Human Potential
Program under contract HPRN-CT-2000-00131 Quantum Space-Time, in
which L. A.,  R. D. and M. A. Ll. are associated to Torino
University. This work has also  been supported by the
D.O.E. grant DE-FG03-91ER40662, Task C.

\end{document}